# Change points in the spread of COVID-19 question the effectiveness of nonpharmaceutical interventions in Germany

Author:
Thomas Wieland
Karlsruhe Institute of Technology, Institute of Geography and Geoecology, Kaiserstr. 12, 76131 Karlsruhe, Germany, E-mail: thomas.wieland@kit.edu.
(Corresponding author)

**Abstract**

**Aims:** Nonpharmaceutical interventions against the spread of SARS-CoV-2 in Germany included the cancellation of mass events (from March 8), closures of schools and child day care facilities (from March 16) as well as a "lockdown" (from March 23). This study attempts to assess the effectiveness of these interventions in terms of revealing their impact on infections over time.

**Methods:** Dates of infections were estimated from official German case data by incorporating the incubation period and an empirical reporting delay. Exponential growth models for infections and reproduction numbers were estimated and investigated with respect to change points in the time series.

**Results:** A significant decline of daily and cumulative infections as well as reproduction numbers is found at March 8 (CI [7, 9]), March 10 (CI [9, 11] and March 3 (CI [2, 4]), respectively. Further declines and stabilizations are found in the end of March. There is also a change point in new infections at April 19 (CI [18, 20]), but daily infections still show a negative growth. From March 19 (CI [18, 20]), the reproduction numbers fluctuate on a level below one.

**Conclusions:** The decline of infections in early March 2020 can be attributed to relatively small interventions and voluntary behavioural changes. Additional effects of later interventions cannot be detected clearly. Liberalizations of measures did not induce a re-increase of infections. Thus, the effectiveness of most German interventions remains questionable. Moreover, assessing of interventions is impeded by the estimation of true infection dates and the influence of test volume.

**Word count (excluding the abstract, references, figures and tables)**: 4143

**Background**

Assessing the effectiveness of nonpharmaceutical interventions (NPIs) in the SARS-CoV-2/COVID-19 context is a topic of growing relevance. Nevertheless, findings documenting the impact of these measures have not been homogeneous within the literature; whether with respect to single countries [1-11], or in terms of international comparisons [12-17]. The question of whether "lockdowns" – including contact bans, curfews or closures of schools and child day care facilities – succeed or fail in reducing infections is a key concern for policymakers, as such measures are accompanied by consequences in terms of economic, social and psychological effects on societies. All European countries introduced NPIs to reduce infections, ranging from appeals to voluntary behaviour changes and the cancellation of mass events (Sweden) to strict curfews (e.g. France, Italy). Being one of the most affected countries (in terms of confirmed prevalence), Germany introduced a strict strategy incorporating three bundles of measures (1. cancellation of mass events after March 8, 2. closure of schools and child day care facilities between March 16 and 18, and 3. a contact ban, bans of gatherings and closures of "nonessential" services from March 23).

There have been some approaches to assessing the interventions in Germany: Dehning et al. [1] utilized epidemiological models (the SIR [susceptible-infected-recovered] model and its extensions) combined with Bayesian inference to find change points in infections over time with respect to the aforementioned measures. They identified impacts of all three bundles of interventions and on this basis have explicitly outlined the importance and necessity of the contact ban for reducing new infections. In a series of studies [2-5], German economists investigated structural breaks in time series of cumulated infections and growth rates. Their inferred change points have been interpreted in a similar way, i.e., in support of the measures. An additional modelling approach using a modified SIR model [4] also outlines the impact of NPIs on infections.

The common denominator in the approaches mentioned above [1-5] is the application of disease case data from the Johns Hopkins University (JHU). This data differs from the official German case data provided by the Robert Koch Institute (RKI) in terms of both precision and detail, with importantly, the latter dataset including information about the date of onset of symptoms for most cases [18,19]. This information is essential because it helps to estimate the true infection dates. In the aforementioned studies [1-5], information of this type was not available, which has therefore required assumptions to be formulated regarding the time between infection and reporting. The SIR modeling study [1] has already been criticized in terms of its underestimation of this delay and the related results [20]. Moreover, studies utilizing epidemiological models [1,4,6] require assumptions on the transmission process of the disease (e.g., spreading rate, contacts per capita) or other unknown epidemiological parameters. Both aspects raise the question whether the previous assessments of NPIs in Germany are reliable.

**Aims**

The aim of this study is to assess the effectiveness of NPIs towards the SARS-CoV-2 spread in Germany (from March 8, 16 and 23, respectively), while overcoming the data-related problems mentioned above. The measures are analysed in terms of revealing their impact on infections over time. By using official case data [19], true dates of infection are estimated. Inspired by the methodical approach in previous studies [2-4,11], change points in time series of three indicators (daily and cumulative infections as well as reproduction numbers, all of which were calculated based on the estimated infection dates) were detected. The data covers infections from February 15 to May 31, 2020, which means that also possible effects of the easing of measures (from April 20) and the introduction of face masks (from April 27) can be assessed.

**Estimating the dates of infection**

To assess the effectiveness of NPIs, it is the dates of infections of the reported cases which must be regarded, rather than the date of report. However, the real time of infection is unknown, thus, it must be estimated using the reported cases. In simple terms, the time between infection and reporting consists of two time periods: a) the time between infection and onset of symptoms (incubation period), and b) the time between onset of symptoms and the date of report (reporting delay). Thus, to estimate the date of infection, both periods must be subtracted from the date of report [1-5,8,11].

There are several estimations of the SARS-CoV-2/COVID-19 incubation period, ranging from median values of 5.0 to 6.4 days [21,22]. Incorporating the reporting delay, however, is much more difficult. Previous studies investigating the effectiveness of interventions in Germany [1-5] have employed data from the Johns Hopkins University (JHU) which only includes daily infection and death cases. The reporting delay is either assumed to be equal to 2-3 days [2-4] or estimated in the model parametrization [1,5]. In contrast, the data on German cases from the Robert Koch Institute (RKI) includes the reporting date and, for the majority of cases, case-specific dates of onset of symptoms, socio-demographic information (age group, gender), and the corresponding county [18,19]. The data used here is the RKI dataset from June 28, 2020 [19]. In this dataset, there were 193,467 reported infections, for which, the date of onset of symptoms is known in 135,967 cases (70.28 %). The arithmetic mean of the time between onset of symptoms and report (reporting delay) is equal to 6.71 days (SD = 6.19) and the corresponding median equals 5 days. 95 % of the reporting delays lie between 0 (2.5 % percentile) and 21 (97.5 percentile) days. On this basis, we clearly see that assuming this value to be equal to 2-3 days [2-4] is an obvious underestimation. Moreover, exploring the dataset reveals that the reporting delay varies between the age and gender groups of the reported cases and over time, as well as between German counties. These differences indicate that it is difficult to assume or estimate average values for the reporting delay [1,5].

Thus, the estimation of the true infection dates of reported cases was conducted using the information from the RKI case data. In line with previous studies [1-5], the incubation period is assumed to equal 5 days, which is the minimum value reported in the literature [21,22]. Given this time period for the records in the case dataset with known date of onset of symptoms, the date of infection of case $i$, $DI_i$, is calculated as the date of onset of symptoms ($DO_i$) subtracted by the incubation period ($IP$):

$$DI_i = DO_i - IP$$

Based on the cases with full information, a dummy variable regression model was estimated for the interpolation of the reporting delay for the remaining 57,500 cases. As the reporting delay differs across case-specific attributes, the reporting delay for case $i$ ($RD_{i,agcwt}$) was estimated by including dummy variables for age group $a$ ($a = 1, …, A$), gender group $g$ ($g = 1, …, G$), county $c$ ($c = 1, …, C$) and weekday $w$ ($w = 1, …, W$) as well as the time trend $t$:

$$RD_{i,agcwt} = \alpha + \sum_{a}^{A-1} \beta_a D_{agegroup_a} + \sum_{g}^{G-1} \gamma_g D_{gender_g} + \sum_{c}^{C-1} \delta_c D_{county_c} + \sum_{w}^{W-1} \zeta_w D_{weekday_w} + \varphi\, t + \varepsilon_{i,agcwt}$$

where $\alpha$ is the estimated constant, $\beta_a$, $\gamma_g$, $\delta_c$ and $\zeta_w$ represent sets of empirically estimated parameters for the *A-1* age groups, *G-1* gender groups, *C-1* counties and *W-1* weekdays, $\varphi$ is the empirically estimated parameter for the time trend and $\varepsilon_{i,agcwt}$ is the stochastic disturbance term. The model parametrization was conducted via Ordinary Least Squares (OLS) estimation.

In those cases lacking the information on onset of symptoms, the date of infection was calculated as the date of report ($DR_i$) subtracted by the estimated reporting delay and the incubation period:

$$DI_i = DR_i - RD_{i,agcwt} - IP$$

**Infection indicators and detection of change points over time**

Previous studies with respect to the assessment of interventions have focused on only one indicator such as daily infections [1], cumulative infections [2-5,8,11,15], or reproduction numbers [7,17]. To arrive at a more holistic picture, three indicators are used: a) the daily new infections, b) cumulative infections and c) the daily reproduction numbers. The estimated infections dates ($DI_i$) were summarized over days which results in the daily new infections at time $t$ ($I^D_t$) and the corresponding cumulative infections at time $t$ ($I^C_t$). The reproduction number for time $t$ ($R_t$) was computed according to the calculation provided by the Robert Koch Institute [18] as the quotient of infections in two succeeding 4-day intervals (implying a generation period of 4 days):

$$R_t = \frac{\sum_{t-4}^{t-3} I^D_t}{\sum_{t-7}^{t-4} I^D_t}$$

The period under study includes the infections from February 15 (first proven "super spreading event" in Germany, the "Kappensitzung" in Gangelt, North Rhine Westphalia) to May 31, resulting in N = 107 daily observations. The final date is estimated by the last available date of report (June 27) subtracted by the 97.5 % percentile of the incubation period (5.6 days) and the 97.5 % percentile of the reporting delay (21 days).

For the analysis of infections over time, phenomenological models have the advantage that they only incorporate time series of infections and do not require further assumptions concerning the transmission process of the disease under study [23,24]. Thus, the time series of all three indicators were analysed using exponential growth models in their semilog form, which means that the dependent variables ($I^D_t$, $I^C_t$ and $R_t$) were transformed via natural logarithm. The model parametrization was conducted via Ordinary Least Squares (OLS) estimation. The corresponding slope parameter of the independent variable (time), here denoted as $\lambda$, represent the average growth rate per time unit (days) and $\lambda$*100 equals the percentage change per day:

$$\ln(I^D_t) = \alpha^D + \lambda^D\, t + \mu^D_t$$

$$\ln(I^C_t) = \alpha^C + \lambda^C\, t + \mu^C_t$$

$$\ln(R_t) = \alpha^R + \lambda^R\, t + \mu^R_t$$

where $\alpha^D$, $\alpha^C$, $\alpha^R$, $\lambda^D$, $\lambda^C$ and $\lambda^R$ are the parameters to be estimated and $\mu^D_t$, $\mu^C_t$ and $\mu^R_t$ represent the stochastic disturbance term in each model.

The detection and dating of change points was conducted using a fluctuation test (recursive estimation test) and *F* statistics, which incorporates comparing the regression coefficients of a time series with *M* breakpoints (and, thus, *M*+1 segments) to the full sample estimates (no segmentation). Within these tests, structural breaks in the time series can be identified. The optimal number of breakpoints and their attribution to the specific observation at which point they occur (which means a dating of the breakpoint, including the computation of confidence intervals) was conducted using the Bai-Perron algorithm. The statistically optimal number of *M* breakpoints is inferred by comparing model variants with zero to five breakpoints (corresponding to one to six segments). The variant which minimizes the residual sum of squares (RSS) and the Bayesian information criterion (BIC) is considered to be the optimal solution [25,26]. Thus, the exponential growth functions shown above are divided into *M*+1 segments, in which the regression coefficients in each *m* segment (*m* = 1, …, *M*+1) are constant. The analysis was conducted in *R* [27] using the package *strucchange* [26].

**Results**

Fig. 1 shows the daily reported cases in the RKI dataset, the corresponding daily onsets of symptoms (incorporating the reporting delay) and the daily infections (incorporating the reporting delay and the incubation period) from February 15 to May 31, 2020. Fig. 2 presents the estimated infections and reported cases on the level of calendar weeks along with additional information about the number of conducted SARS-CoV-2 tests [28]. Obviously, the time series are not simply shifted by the average delay between infection and report. The differences between the temporal development of infection and report curves can be attributed to temporal, case-specific, and regional differences in the reporting delay. Furthermore, all results emerging from time series of infections shown below have to be interpreted whilst taking into consideration the changing number of tests conducted weekly. Specifically, we can see an increase in the number of tests by a factor of 2.73 from calendar week 11 (127,457 tests) to 12 (348,619 tests), followed by smaller fluctuations in the succeeding weeks.

Fig. 3, 4 and 5 show the results of the time series analysis of daily infections, cumulative infections, and reproduction numbers, respectively. The top-left plot shows the optimal structural breaks in time series and the corresponding slopes (exponential growth rates) for each model segment. The top-right plot displays the explained variance ($R^2$) and the point estimate confidence intervals for each model segment. The bottom-left plot presents the corresponding model diagnostics (BIC and RSS) for the model variants with one to five segments, and the adjacent plot shows the model fit on condition that no structural breaks occur. With respect to daily infections (fig. 3), the best model fit minimizing BIC and RSS incorporates three breakpoints and four model segments, respectively. Obviously, a model without breakpoints does not fit the time series appropriately. The three significant structural breaks are on March 8 (95 % confidence intervals: March 7 to March 9) and March 24 (CI [23, 25]) as well as April 19 (CI [18, 20]). The first break on March 8 reduces the growth rate from 0.229 (CI [0.217, 0.240]), which represents an average daily increase of 22.9 % (February 15 to March 8), to -0.013 (CI [-0.025, -0.001]), which means a daily decrease equal to 1.3 % (March 9 to March 24). From March 25, the daily infections decrease by 5.4 % per day (-0.054, CI [-0.057, -0.050]) until April 19. From April 20, the decline of new infections slows down but the daily growth rate is still negative with -3.0 % (-0.030, CI [-0.033, -0.028]).

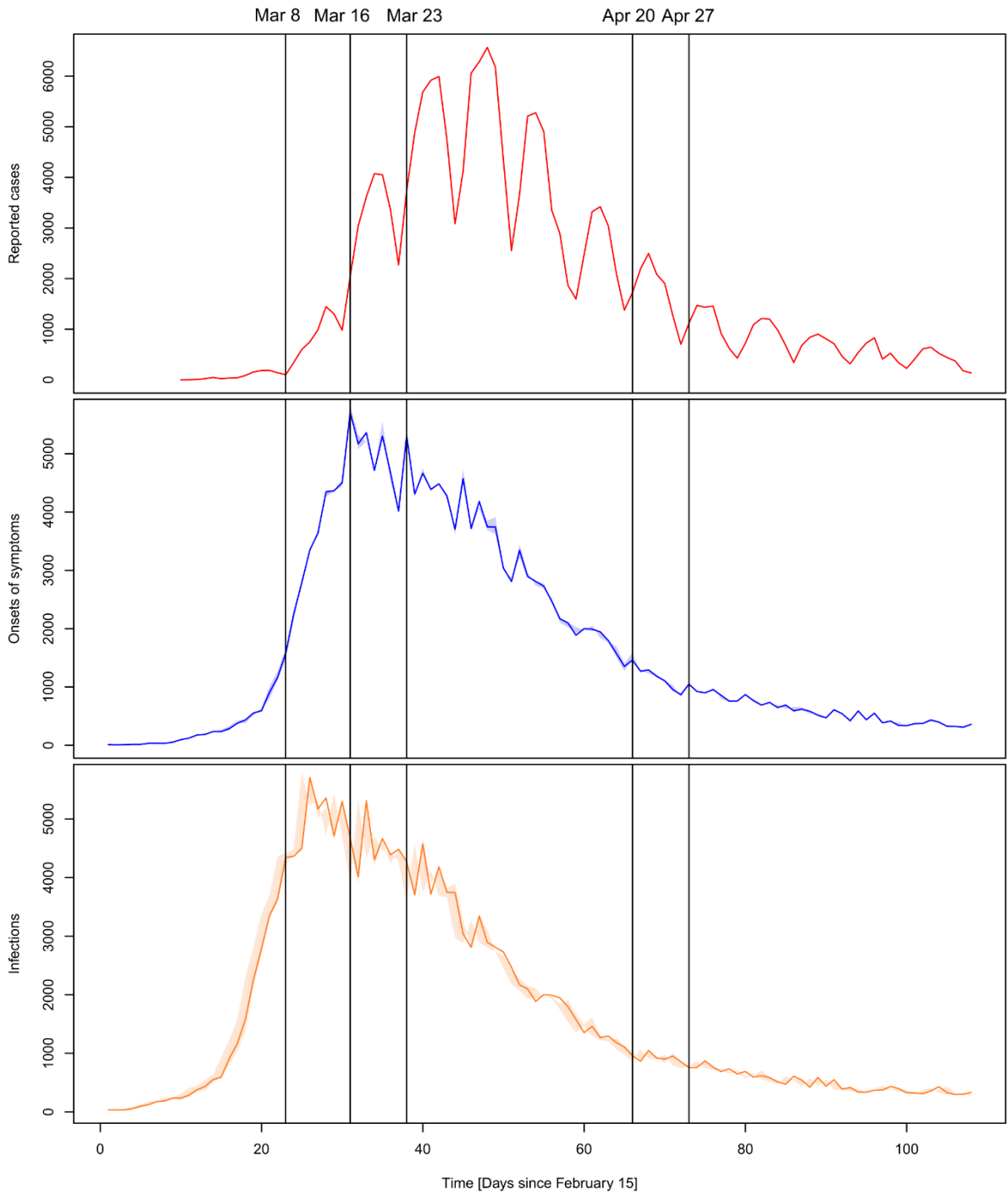

Figure 1: Daily values of reported cases, onsets of symptoms and infections from 15 February to 31 May 2020.

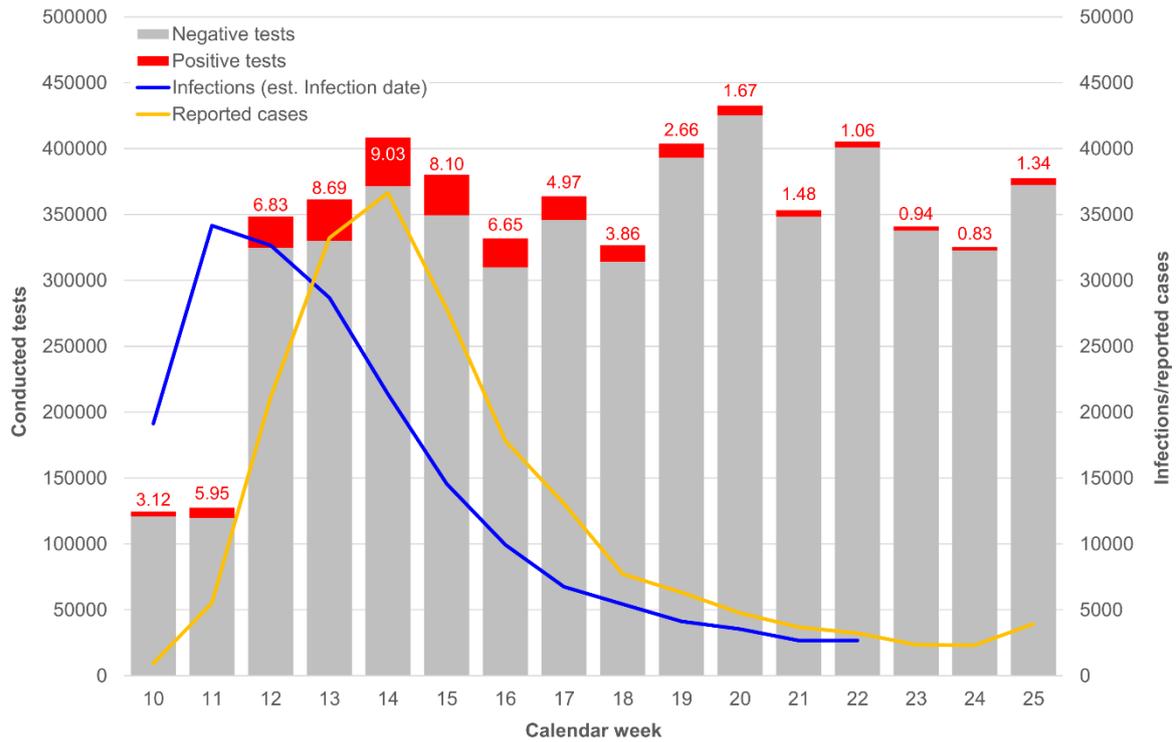

Figure 2: Weekly values of reported cases, infections and conducted SARS-CoV-2 tests from calendar week 10 to 25.

The best model solution for the cumulative infections over time also incorporates three breakpoints. The first break occurs on March 10 (CI [9, 11]), at which point the daily growth rate was reduced from 22.8 % (0.228, CI [0.224, 0.232]) to 6.6 % (0.066, CI [0.059, 0.073]). The second break on March 26 (CI [25, 27]) documents a further decrease in daily growth from 6.8 % to 1.9 % (0.019, CI [0.017, 0.020]). The last structural change is detected on April 13 (CI [12, 14]), at which time the daily growth rate shifted from 1.9 % to 0.4 % (0.004, CI [0.003, 0.004]).

With respect to the reproduction number (R), three structural breaks can also be identified. After the first break on March 3 (CI [2, 4]), R starts to decrease by 9.7 % per day (-0.097, CI [-0.107, -0.087]) until March 19 (CI [18, 20]). The break around March 19 initiates a stabilization of the R value with a decrease equal to 0.7 % per day (-0.007, CI [-0,008, -0,005]). From the last change point which occurs at April 23 (CI [22, 27]), the reproduction number still fluctuates on a low level with a daily increase of 0.3 % (0.003, CI [0.000, 0.005]). With few exceptions, from March 19, the daily reproduction number remains below one (ln R < 0).

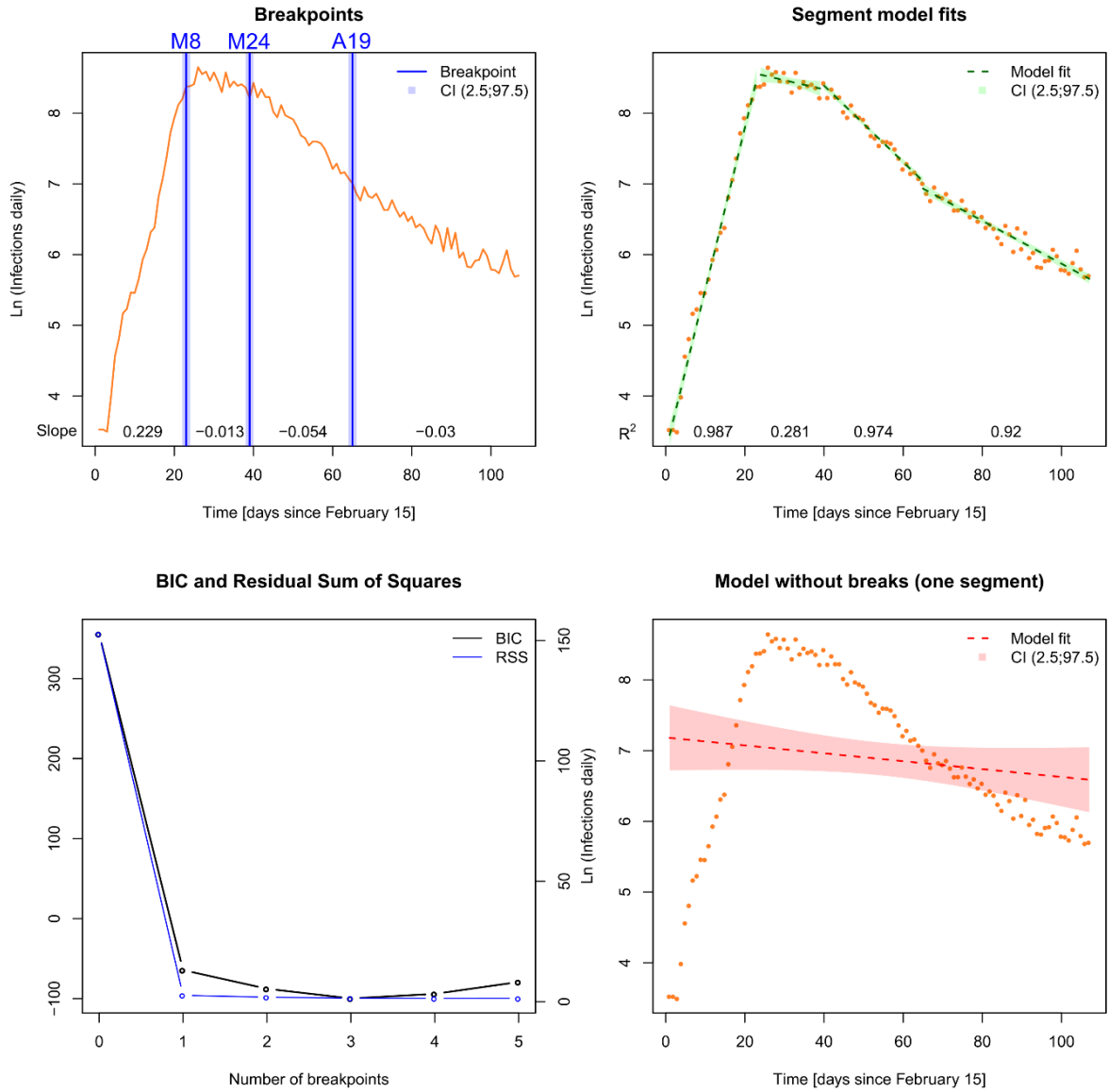

Figure 3: Time series and corresponding break points as well as model diagnostics for daily infections from 15 February to 31 May 2020.

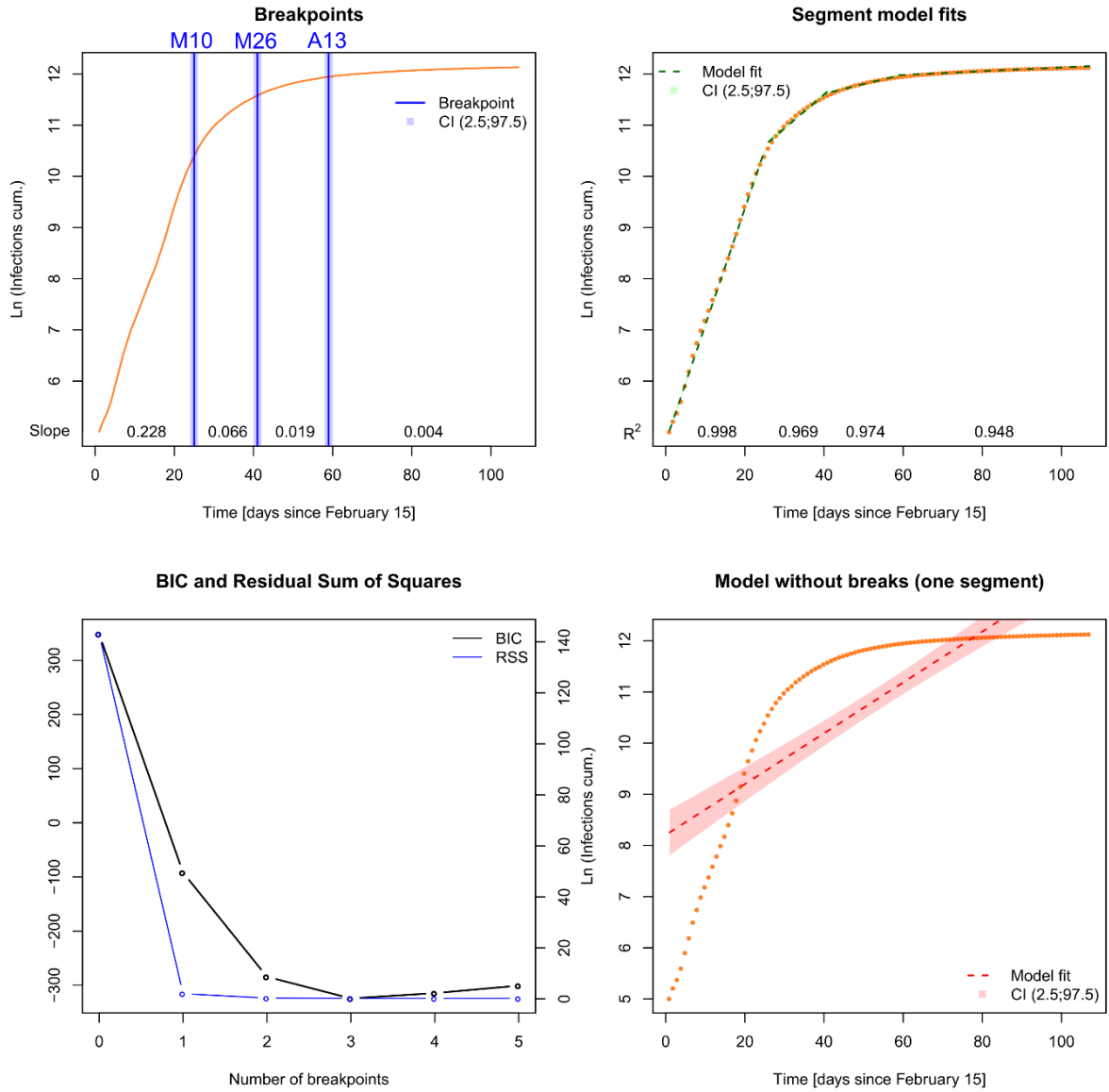

Figure 4: Time series and corresponding break points as well as model diagnostics for cumulative infections from 15 February to 31 May 2020.

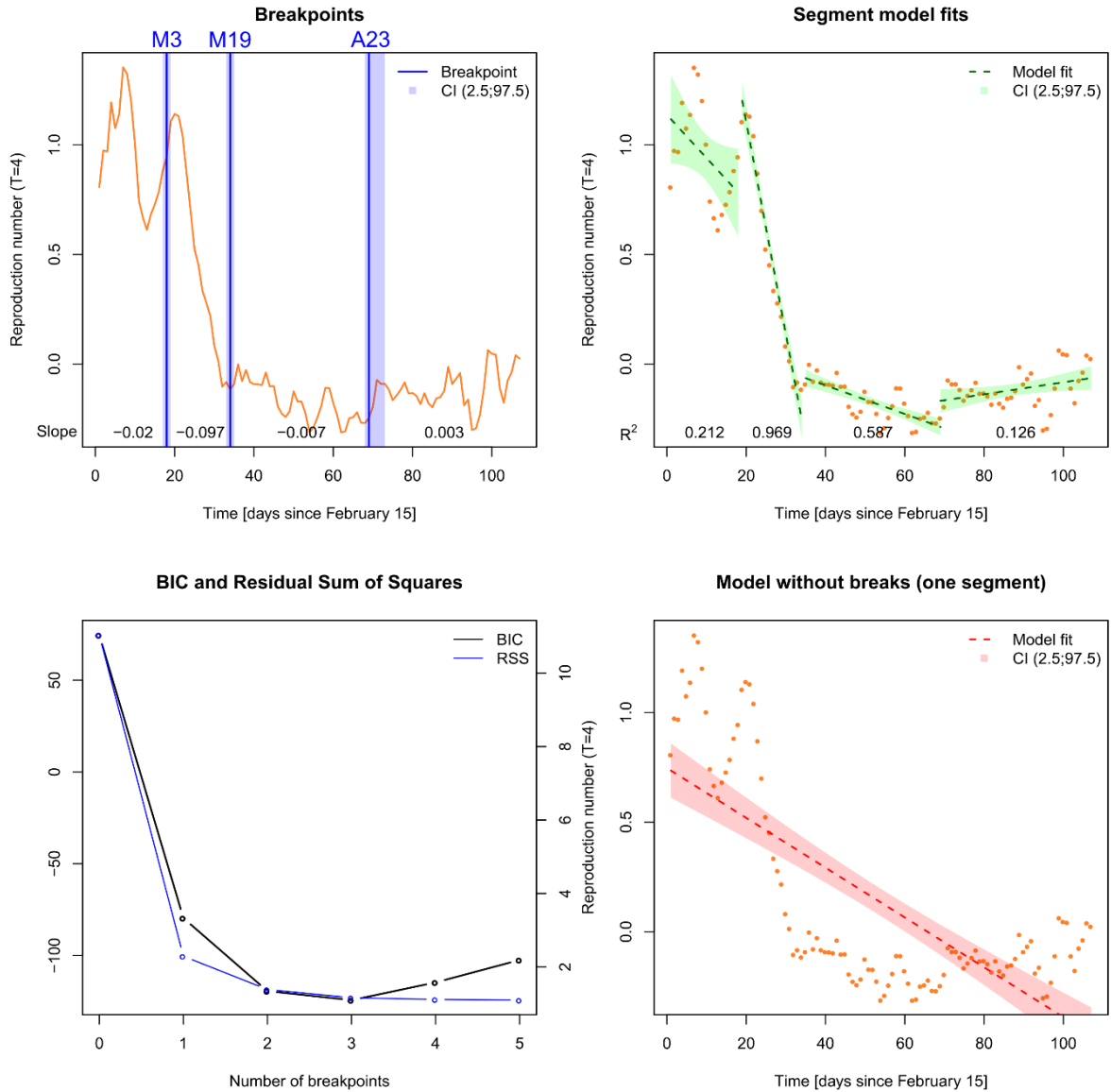

Figure 5: Time series and corresponding break points as well as model diagnostics for reproduction numbers (R) from 15 February to 31 May 2020.

All in all, we find concordant structural breaks for all three indicators in the first third of March 2020. Around March 8, the daily new infections turn from exponential growth to decay and the growth rate of cumulative infections has its highest decrease. This decline occurs although the test volume increased strongly in the succeeding weeks (see fig. 2). Unfortunately, conducted tests cannot be linked to reported cases as both information stem from different data sources. However, the massive increase of testing must have had an influence on the detection of SARS-CoV-2 infections occurred before. It is therefore plausible to assume that if test volume had remained constant over time, fewer infections would have been detected and the decrease of (confirmed) infections would have been stronger. The other breakpoints are not coincident: Whilst structural changes in the daily and cumulative infections occur in the last third of March, there is no corresponding break with respect to reproduction numbers. In the last third of April, we find structural changes with respect to daily infections and reproduction numbers, but the growth rate of infections still remains negative. From March 19, the reproduction numbers, with few exceptions, fluctuate on a level below one (ln R < 0).

**Discussion**

Regarding all three indicators, we find consistent results with respect to a significant decline of infections in the first third of March – about one week before the closing of schools and child day care and two weeks before the full "lockdown" (including the contact ban) came into force. The effect coincides with the cancellation of mass events recommended by the German minister of health, Spahn, on March 8. However, the increased awareness in the general population could have also had a significant impact in terms of voluntary changes in daily behaviour (e.g., physical distancing to strangers, careful coughing and sneezing, thorough and frequent hand washing). Surveys demonstrate an increased awareness towards the Corona threat already in the middle of February [29]. Additionally, voluntary cautious behaviour in the Corona context could also explain the abrupt and unusual decline of *other* infectious respiratory diseases in Germany starting in early March [30].

Previous studies have also found a first slowing of infections in the first third of march [1-5], but the results of the present analysis contradict their findings as the change point in the 10th calendar week is a) the clearest structural break given that it is present for all three indicators, b) the break which initiated a trend change in terms of a decline of daily new infections and c) the most influential break with respect to cumulative infections. Dehning et al. [1] state: "Our results indicate that the full extent of interventions was necessary to stop exponential growth […] Only with the third intervention, the contact ban, we found that the epidemic changed from growth to decay". These statements are based on a negative growth rate (-3 %) having not become apparent before the contact ban came into force. In contrast, given the estimated infection dates in the present study, we see that the growth rates of new infections and reproduction numbers already turn negative on March 3 and 8, respectively. At the same time, the growth rate of the cumulative infections has its biggest decrease across all four segments of the time series. Thus, a decline in infections occurred before school closures and the contact ban came into force.

In the time series studies on the German case [2-5], the closing of infrastructures (schools etc.) in mid-March was found to be the most influential break with respect to cumulative infections. This conclusion cannot be confirmed in the present study, as we cannot find any referring breakpoint with respect to the daily and cumulative infections. If the closures of schools and child day care facilities would have had an impact on infections, there would have been a significant decline of new infections from March 16 to 18 on. The structural break in the reproduction numbers on March 19 initiates a stabilization of the reproduction numbers but not a further decline. Therefore, an impact of school and child day care closures cannot be detected. The influence of the third intervention ("lockdown" including contact ban), which was found to be the most influent factor in the SIR modelling study [1], and an important factor in the previous time series analyses [2-5], remains unclear in the present study as well. There is no structural break in the reproduction numbers which coincides with the contact ban. Significant breaks in daily and cumulative infections occur after the social ban came into force, but not immediately. The mismatches between the present and previous results are obviously related to different data sources, a point underscored by the fact that the modelling approach is similar to some of the previous studies [2-4].

The impact of first liberalizations of measures from April 20 (e.g., reopening of some "nonessential" retail shops) is plausibly reflected in the temporal development of new infections and reproduction numbers. However, there is no re-increase of new infections as the corresponding growth rate remains negative and the reproduction numbers remain, with few exceptions, below the critical value of one. Moreover, no effect of the implementation of compulsory face masks in retail shops and public transport (starting from April 27) can be detected, as there is no further significant structural break. However, this intervention was implemented at a time where infections were already on a low level. Thus, the effectiveness of this measure cannot be definitely assessed. Further liberalizations starting in the first half of May (e.g., reopening of schools for some age groups, extending emergency childcare) do not show any impact as well.

The current findings support results for Germany inferred from logistic growth models which show a trend change before the contact ban came into force [8]. In addition, a Spanish time series study revealed breakpoints in cumulative infections, with the first occurring about two weeks before the nationwide "lockdown" [11]. Furthermore, the present results tend to support other studies of international comparisons which have found a decline of infections with or without strict interventions [12-16].

**Strengths and limitations**

One strength of the present study is the relative simplicity of the analysis. The current approach allows for a time-related analysis of NPIs based on a rather simple model which does not require further assumptions concerning the disease under study. Thus, the methodology can be easily transferred to other pandemics, countries, or regions as only time series of infections are necessary. In the future, the research design should be applied to international comparisons, incorporating both Scandinavian and South-European countries. Another strength is the utilization of realistic infection dates, which was not incorporated in previous studies. Moreover, regarding three different indicators allows for a more differentiated picture of infections over time.

The temporal development of the three indicators was also contrasted with conducted tests over time. However, in the absence of daily test data, the impact of changing test volumes was not assessed directly. Another limitation results from the phenomenological nature of the regression models utilized for time series analysis. As the only explanatory variable is time, we can question the impacts of the regarded interventions but cannot explain the factors causing the temporal development of infections directly.

**Conclusions**

This study finds clear evidence of a decline of SARS-CoV-2 infections in Germany at the beginning of March 2020, which can be attributed to relatively small nonpharmaceutical interventions (cancellation of mass events) and voluntary behavioral changes. A trend change of infections from exponential growth to decay was not induced by the "lockdown" measures but occurred earlier. Additional impacts of later NPIs cannot be clearly detected: Firstly, there is no significant effect with respect to infections that could be attributed to school and day-care closures. Secondly, effects which could be related to the contact ban a) do not appear with respect to all three indicators, b) differ in strength and tend towards lower impacts, and c) do not match the time the measure came into force. Thus, the necessity of the second (March 16-18) and the third bundle of interventions (March 23) is questionable because a) the related effects on infections (if any) cannot be unequivocally validated, b) a trend change had already occurred long before they came into force, and c) liberalizations of these measures did not induce a re-increase of infections. We cannot deduce conclusions towards the necessity of compulsory face masks, as this intervention was introduced late. Furthermore, the time series of (confirmed) infections is substantially influenced by temporal changes in the test volume, which leads to a high degree of uncertainty with respect to the data source. Therefore, a future evaluation of NPIs towards SARS-CoV-2/COVID-19 in Germany should consider these questionable effects and uncertainties.

The study reveals three methodological issues for assessing the impact of NPIs which may influence the results enormously. Firstly, the key challenge is the estimation of realistic infections dates from official statistics (which typically do not include this information). This information is essential for the assessment of measures which aim at the reduction of new infections. It is particularly important to include a realistic and differentiated reporting delay. An underestimation of the time between infection and reporting leads to the estimation of infections to a later date than actually occurred in reality. As a consequence, trend changes will also be dated too late, and thus, are attributed erroneously to specific interventions. Secondly, it is important to incorporate several indicators for the pandemic spread. Daily and cumulative infections as well as reproduction numbers, though based on the same initial data, have different meanings. As the results of this study show, significant change points may be found for some indicators but not for others. Thus, assessment of effectiveness of nonpharmaceutical interventions depends on the indicator used which leads to the conclusion that the temporal development of the indicators chosen should be carefully compared. And lastly, quantitative investigations based on empirical case data implicitly assume constant test volumes, which is obviously not true. In the German case, the number of conducted tests for SARS-CoV-2 is not constant over time. An increase (or decrease) of tests may result in an artificial increase (or decrease) of reported infections. Thus, increasing test capacity – which is a key parameter in fighting a pandemic – may result in a statistical source of error when analyzing pandemics over time. All these issues exist regardless of the chosen modeling approach, which suggests a need to shift study design toward prioritizing the handling of data sources rather than refining models.

**Conflict of interest**
The author declares that there is no conflict of interest.

**Funding**
The author received no specific funding for this work.